\documentclass[english,aps,twocolumn,superscriptaddress,jap]{revtex4-1}
\usepackage[T1]{fontenc}
\usepackage[latin9]{inputenc}
\usepackage{units}
\usepackage{textcomp}
\usepackage{graphicx}
\usepackage{amsmath}

\DeclareGraphicsExtensions{.pdf,.png,.jpg}

\begin{document}

\title{Computation and visualization of accessible reciprocal space and resolution element in high-resolution X-ray diffraction mapping}

\author{Tatjana Ulyanenkova}

\address{Rigaku Europe SE, Am Hardwald 11, 76275 Ettlingen, Germany}

\author{Aliaksei~Zhylik}

\address{Atomicus OOO, Mogilevskaya Str.\ 39a-530, 220007 Minsk, Belarus}

\author{Andrei Benediktovitch}

\address{Atomicus OOO, Mogilevskaya Str.\ 39a-530, 220007 Minsk, Belarus}

\author{Alex Ulyanenkov}

\address{Atomicus GmbH, Schoemperlen Str.\ 12a, 76185 Karlsruhe, Germany}

\begin{abstract}
An approach to visualize the accessible reciprocal space accounting the goniometer angles limitation and the resolution element in the reciprocal space is presented. The shapes of the accessible reciprocal space region for coplanar and non-coplanar geometries are given employing the additional degree of freedom provided by detector arm. The examples of the resolution elements in different points of the reciprocal space are shown. The equations obtained permit to find these regions and to calculate the experimental geometry to obtain the diffraction from any accessible reflections, which makes a further extension of previously reported methods. The introduced algorithm has been testified by experimental measurements of reciprocal space maps in coplanar and non-coplanar geometries. The examples of the resolution elements in different points of the reciprocal space are calculated to illustrate the method proposed.
\end{abstract}

\maketitle

\section{Introduction}
\label{intro}

The high-resolution X-ray diffraction (HRXRD) experiments for thin film analysis involve quite complicated techniques for geometrical configuration of the experiment and calculations of expected resolution parameters. The planning of such measurements is one of the important stages of the X-ray experiment and data analysis. The properly done planning can reduce the measurement preparation time, as well as bring closer the success of the result. The visualization tool for the accessible reciprocal space area providing the schematic drawing of the sample's Bragg reflections simplifies the planning of X-ray diffraction experiment. Usually the accessible reciprocal space regions are drawn for the case of coplanar diffraction, when the incident, the outgoing wave vectors and the surface normal lie in the same plane, \cite{bowen2005}. There are publications available where the accessible reciprocal space is demonstrated in 3D for both coplanar and non-coplanar diffraction, \cite{yefanov2008} and the appropriate visualization software is developed by \cite{yefanov2008a}. This contribution makes a further extension of the previous works and presents an approach to visualize the accessible reciprocal space in the case when the vertical goniometer with a detector arm possesses two degrees of freedom, which enables the performance of the non-coplanar diffraction without sample inclination. The scattering vector is parameterized by angular variables associated with the physical axes of the goniometer, which takes into account the angular limitations of the goniometer. We propose below the analytical expressions for angular coordinates in considered parametrization for the position of the reciprocal lattice vector, which makes possible to visualize the resolution element in the reciprocal space.

The additional in-plane degree of freedom provided by the detector arm is very beneficial for the analysis of thin films, \cite{2002Ofuji}, \cite{2007yoshida2007x}, for the texture analysis, \cite{2011nagao2011x}, for residual stress gradients investigation \cite{2014BenediktovitchIPStress} and the other applications. In the case of high-resolution X-ray diffraction applications, the advantage of non-coplanar diffraction with two degrees of freedom detector arm is provided by an access to those reflections, which involve the grazing angles in a coplanar mode. The examples of such a geometry are shown below for Ge/Si(001) sample for $(113)$ and $(\overline{1}\overline{1}3)$ reflections, which require the incidence (outgoing) angles in the range of 0 to 5 degrees in coplanar case, and for $(1 \overline{1}3)$ reflection in non-coplanar case, which does not involve grazing angles.

\section{Basic principles}
\label{sec:1}

X-ray diffraction methods are based on X-rays elastic scattering from the atoms of the investigated substance. The incident wave is characterized by the wave vector $\vec{k}_{in}$, and the scattered wave propagated in the detector direction is characterized by the wave vector $\vec{k}_{out}$. Due to an elastic nature of scattering process $\left|\vec{k}_{out}\right|=\left| \vec{k}_{in} \right| = k_0 = \frac{2 \pi}{\lambda}$, where $\lambda$ is a wavelength of the incident and scattered waves and the diffraction vector $\vec{Q} = \vec{k}_{out} - \vec{k}_{in}$ contains the information on the sample structure.
There are three coordinate systems used for the analysis and for the interpretation of X-ray diffraction results: the laboratory coordinate system $\vec{L}_x$, $\vec{L}_y$, $\vec{L}_z$, the sample coordinate system $\vec{S}_x$, $\vec{S}_y$, $\vec{S}_z$ and the crystal coordinate system \cite{ulyan2014}, \cite{zhylik2013covariant}.
The crystal coordinate system is linked to the cell symmetry of the studied crystal and is convenient for the description of planes and directions using the Miller index form. The laboratory coordinate system is associated with the goniometer position in space and, finally, the sample coordinate system is related to the sample position in space. Usually the laboratory and the sample coordinate systems are chosen as orthonormal ones, therefore relations between these systems are described by a rotation matrix $\mathbf{M}^{LS}$. Furthermore, we use the laboratory coordinate system as a base system for calculations.
There are many X-ray diffraction studies based on the reciprocal space scanning using the diffraction vector $\vec{Q}$ to find a proper position of a crystal reflection $\vec{H}$ or to study the vicinity of a reciprocal lattice point. During the measurement process, the positions of a source, detector and a sample vary consistently, and the goniometer axes related to the positioning of a detector and a source  drive the diffraction vector $\vec{Q}$. In the most common case, the pair of angles for the incident and exit beams is enough for their positioning:

\begin{subequations}
\begin{align}
\vec{k}_{in} = k_0 \mathbf{\Theta_{i2}} \mathbf{\Theta_{i1}} \vec{L}_x, \\
\vec{k}_{out} = k_0 \mathbf{\Theta_{o2}} \mathbf{\Theta_{o1}} \vec{L}_x,
\end{align}
\end{subequations}
where $\mathbf{\Theta_{i2}}$, $\mathbf{\Theta_{i1}}$, $\mathbf{\Theta_{o2}}$, $\mathbf{\Theta_{o1}} $ are the rotation operators.
The goniometer axes related to the sample positioning define the matrix $\mathbf{M}^{LS}$.

\section{Accessible reciprocal space}
\label{sec:2}

\subsection{Main formulas for the vertical diffractometer with the additional detector axis}
\label{sec:2a}

\begin{figure}
\caption{The vertical goniometer with a detector arm possesses two degrees of freedom: (a) the axes $L_x$ and $L_z$ of laboratory coordinate system; (b) schematic axes view of the vertical goniometer with a detector arm.}
\label{VerticalGonio}
 \includegraphics[width=0.6\textwidth]{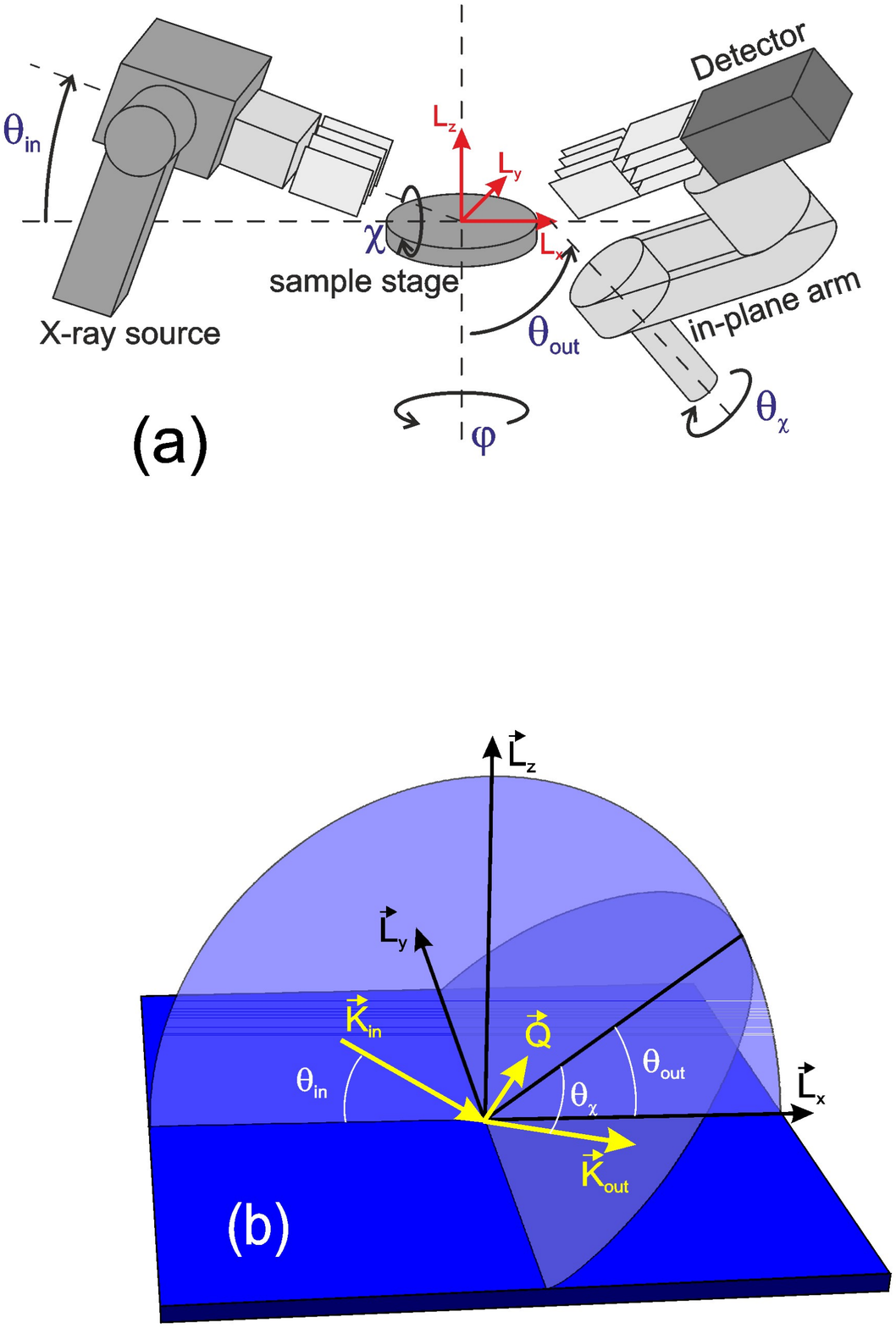}
\end{figure}

The X-ray diffractometer setup (Fig.~\ref{VerticalGonio}(a)) with standard vertical goniometer and additional detector arm is capable to measure X-ray diffraction in coplanar and non-coplanar geometries. As shown on the schematic drawing for the diffraction planes (Figure~\ref{VerticalGonio}(b)), the positions of source and the detector are described by the following angles:
\begin{itemize}
  \item $\theta_{in}$ is the angle between the line connecting sample and X-ray source and the horizontal plane of goniometer;
  \item $\theta_{out}$ is the angle between plane of the axis of in-plane arm rotation and the horizontal plane of goniometer; in case of no in-plane arm rotation $\theta_{out}$ is the angle between the line connecting sample and detector and and the horizontal plane of goniometer;
  \item $\theta_\chi$ is the angle of the in-plane arm rotation.
\end{itemize}

The incidence beam is defined by one rotation angle and for the exit beam two angles with the following order are assigned:

\begin{subequations}
\begin{align}
\vec{k}_{in} = k_0 \mathbf{\Theta_{in}} \vec{L}_x, \\
\vec{k}_{out} = k_0 \mathbf{\Theta_{out}} \mathbf{\Theta_\chi} \vec{L}_x,\\
\mathbf{\Theta_{in}} = \mathbf{R}(\theta_{in}, \vec{L}_y),\\
\mathbf{\Theta_{out}} = \mathbf{R}(-\theta_{out}, \vec{L}_y),\\
\mathbf{\Theta_\chi} = \mathbf{R}(-\theta_\chi, \vec{L}_z),
\end{align}
\end{subequations}
\noindent where
\begin{subequations}
\begin{align}
\mathbf{R}(\theta_{in}, \vec{L}_y) =
\begin{bmatrix}
       \cos(\theta_{in}) & 0 & \sin(\theta_{in}) \\
       0 & 1& 0 \\
       -\sin(\theta_{in}) & 0 & \cos(\theta_{in})
\end{bmatrix},\\
\mathbf{R}(-\theta_{out}, \vec{L}_y) =
\begin{bmatrix}
       \cos(\theta_{out}) & 0 & -\sin(\theta_{out}) \\
       0 & 1& 0 \\
       \sin(\theta_{out})           & 0 & \cos(\theta_{out})
\end{bmatrix},\\
\mathbf{R}(-\theta_\chi, \vec{L}_z) =
\begin{bmatrix}
       \cos(\theta_\chi) &  \sin(\theta_\chi) & 0 \\
       -\sin(\theta_\chi) &  \cos(\theta_\chi) & 0\\
       0 & 0& 1
\end{bmatrix}.\\ \nonumber
\end{align}
\end{subequations}

As a result, the explicit form for the following parameters is obtained:

\begin{subequations}
\label{formulasSurf3D}
\begin{align}
\vec{k}_{in} = k_0 \left\{\cos (\theta_{in}), 0, - \sin (\theta_{in})   \right\}, \\
\vec{k}_{out} = k_0 \left\{\cos (\theta_{out}) \cos (\theta_\chi), -\sin ( \theta_\chi ),
 \sin (\theta_{out})  \cos ( \theta_\chi ) \right\}, \\
\vec{Q} =k_0 \left\{ \cos (\theta_{out}) \cos (\theta_\chi )-\cos (\theta_{in}), -\sin (\theta_\chi ), \right. \nonumber \\
\left.\sin (\theta_{out}) \cos (\theta_\chi
   )+\sin (\theta_{in})  \right\}, \label{formulasQ3D}\\
\left| \vec{Q} \right|^2 = 2 k_0 (1- \cos (\theta_\chi ) \cos (\theta_{out}+\theta_{in})).
\end{align}
\end{subequations}

The diffraction  vector in the laboratory coordinate system is a function of three parameters. The definition range for the  angular parameters $\theta_{in}$, $\theta_{out}$, $\theta_\chi$ may be limited  up to $-\pi \dots \pi $ range without loss of generality.
In the case of goniometer without additional arm for the non-coplanar diffraction geometry ($\theta_\chi \equiv 0$), the equations (\ref{formulasSurf3D}) take the following form:

\begin{subequations}
\label{formulasSurf2D}
\begin{align}
\vec{k}_{in} = k_0 \left\{\cos (\theta_{in}), 0, - \sin (\theta_{in})   \right\},\\
\vec{k}_{out} = k_0 \left\{\cos (\theta_{out}) , 0,  \sin (\theta_{out})   \right\},\\
\vec{Q} =k_0 \left\{ \cos (\theta_{out}) -\cos (\theta_{in}), 0,
\sin (\theta_{out}) +\sin (\theta_{in})  \right\},  \label{formulasQ2D}\\
\left| \vec{Q} \right|^2 = 2 k_0(1 - \cos (\theta_{out}+\theta_{in})).
\end{align}
\end{subequations}

\subsection{Inverse problem}
\label{sec:2b}

The inverse problem consists in the finding the positions $\theta_{in}$, $\theta_{out}$, $\theta_\chi$ for the goniometer axes at certain reciprocal space point for Bragg reflection $\vec{H}$.

\begin{equation}
\frac{\vec{Q}(\theta_{in},\theta_{out}, \theta_\chi )}{k_0} = \vec{H'}.
\end{equation}
where $\vec{H'} = \left(H'_x, H'_y, H'_z \right) = \left( \frac{H_x}{k_0}, \frac{H_y}{k_0}, \frac{H_z}{k_0}\right)$.

 In the case of the goniometer with additional axis for non-coplanar diffraction geometry using the vector $\vec{Q}$ from the equation (\ref{formulasQ3D}), a common solution of the inverse problem contains a set of solutions due to presence of periodical functions, but only 4 of them fall into the definition range of angles $\theta_{in}$, $\theta_{out}$, $\theta_\chi$:

\begin{subequations}
\label{inverce_problev_formula}
\begin{align}
\theta_{\chi 1} = \arcsin(-H'_y) + 2 \pi n, \\
\theta_{\chi 2} = \pi + \arcsin(H'_y) + 2 \pi n,
\end{align}
where $n$ is an integer value.
\begin{align}
\nonumber H_{xz}^{'2} = H_x^{'2} + H_z^{'2}, \\
\nonumber H^{'2} = H_x^{'2} +H_y^{'2} + H_z^{'2}, \\
A = \frac{2H'_z - S \sqrt{4H_{xz}^{'2} - H^{'4}}}{H^{'2} - 2H'_x}, \\
\nonumber S = \pm 1, \\
\theta_{in} = 2 \arctan \left( A \right) + 2 \pi n, \\
\theta_{out} = \arctan \left(\frac{H'_z - \sin (\theta_{in})}{H'_x + \cos (\theta_{in})}\right) + 2 \pi n.
\end{align}
\end{subequations}

\subsection{Available reciprocal space plot}
\label{sec:2c}

\begin{table}
\caption{The angular ranges of a typical goniometer axes for X-ray diffractometer.}
\label{tableLimits}
\begin{tabular}{lcc}      
 axis    & min, $^\circ$        & max, $^\circ$            \\
\hline
 $\theta_{in}$      & -5      & 95            \\
 $\theta_{out}$      & -5      & 120            \\
 $\theta_{\chi}$      & -3      & 120           \\
\end{tabular}
\end{table}

The accessible reciprocal space view is usually connected with a view of a set of available for investigated sample Bragg reflections. Here the accessible reciprocal space will be considered  separately from a set of Bragg reflections and from the sample orientations in the laboratory system. The main goal of such separation is to reduce a number of parameters for reciprocal space representation and to take into account the physical limitations  of the goniometer, which are caused by the instrument design. For the considered in this paper typical goniometer, the ranges of the accessible angles are shown in the Table~\ref{tableLimits}.

\begin{figure}
\caption{Accessible reciprocal space at different angle ranges $\theta_{in}$, $\theta_{out}$ for goniometer without axis for non-coplanar measurements.}
\label{surf2D}
\begin{tabular}{l}
\includegraphics[width=0.7\columnwidth]{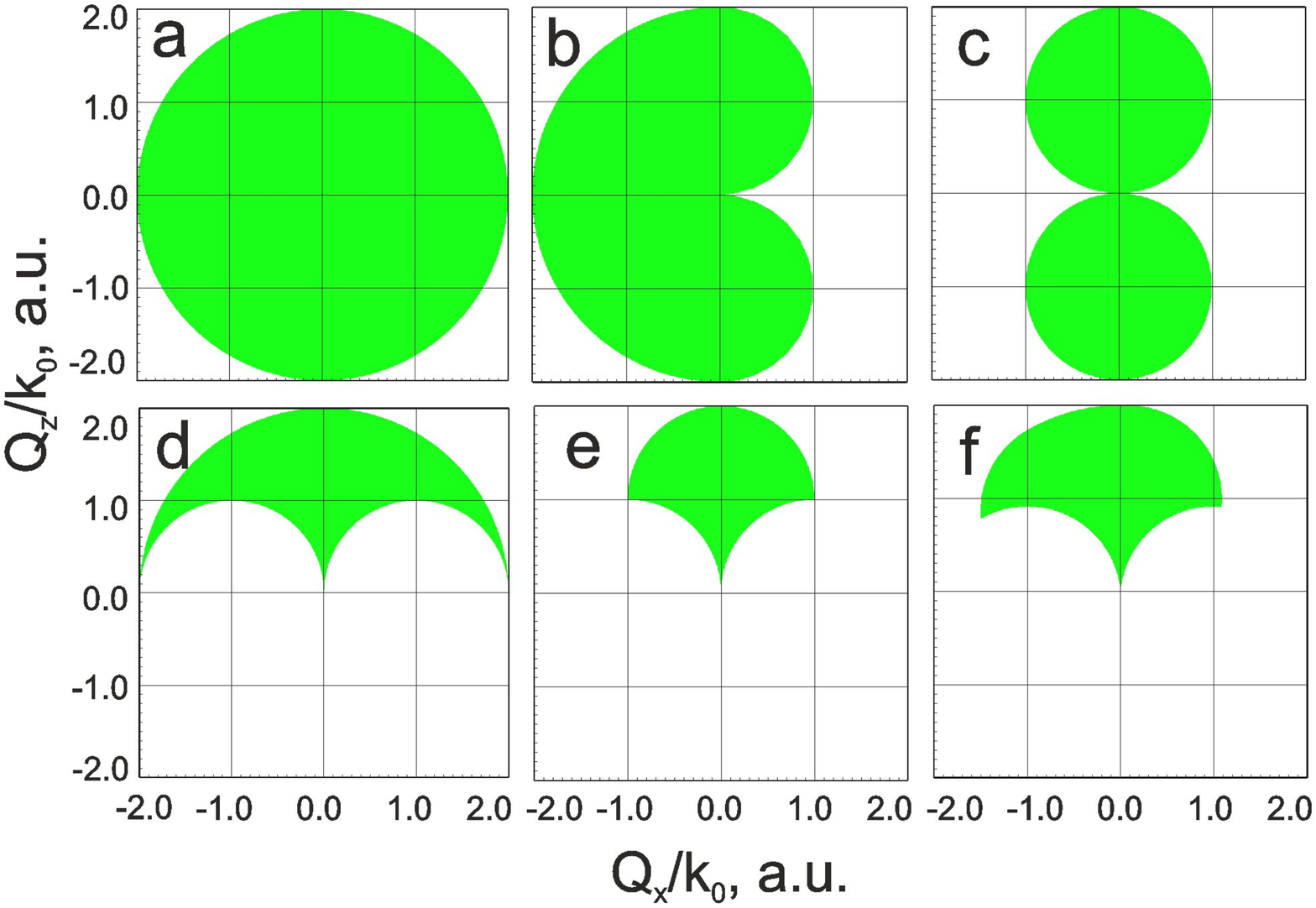} \\
\begin{tabular}{r | c | c | c | c | c | c }
& a & b & c & d & e & f \\ \hline
$\theta_{in, min}$ & $-180^\circ$ & $-90^\circ$ & $-90^\circ$ & $0^\circ$ & $0^\circ$ & $-5^\circ$\\
$\theta_{in, max}$ & $180^\circ$ & $90^\circ$ & $90^\circ$& $180^\circ$ & $90^\circ$ & $95^\circ$\\
$\theta_{out, min}$ & $-180^\circ$ & $-180^\circ$ & $-90^\circ$ & $0^\circ$ & $0^\circ$ & $-5^\circ$ \\
$\theta_{out, max}$ & $180^\circ$ & $180^\circ$ & $90^\circ$ & $180^\circ$ & $90^\circ$ & $120^\circ$\\
\end{tabular}\\
\end{tabular}
\end{figure}

In the case of goniometer without additional axis $\theta_\chi$, all measurements will be done in a coplanar geometry. According to the equation (\ref{formulasQ2D}), the accessible reciprocal space is restricted by the plane $L_x L_z$. Figure~\ref{surf2D} demonstrates the accessible reciprocal space regions for different angle ranges of $\theta_{in}$, $\theta_{out}$.

\begin{figure*}
\caption{Conversion of angular space with $\left\{ \theta_{in} (-\pi, \pi),\theta_{out} (-\pi, \pi), \theta_\chi(-\pi,\pi) \right\}$ ranges to reciprocal space with all component surfaces. Component surfaces at different faces of angular space: a)  $\theta_{in} = -\pi$, b) $\theta_\chi = \pi$, c)  $\theta_{in} + \theta_{out}= -\pi$, d) $\theta_{in} + \theta_{out}= 0$, e) $\theta_{out}= -\pi$, f) $\theta_{in} = \pi$, g) $\theta_\chi = -\pi$, h) $\theta_{in} + \theta_{out}= \pi$, i) $\theta_{out}= \pi$.}
\label{Surf3DFull}
\includegraphics[width=0.8\textwidth]{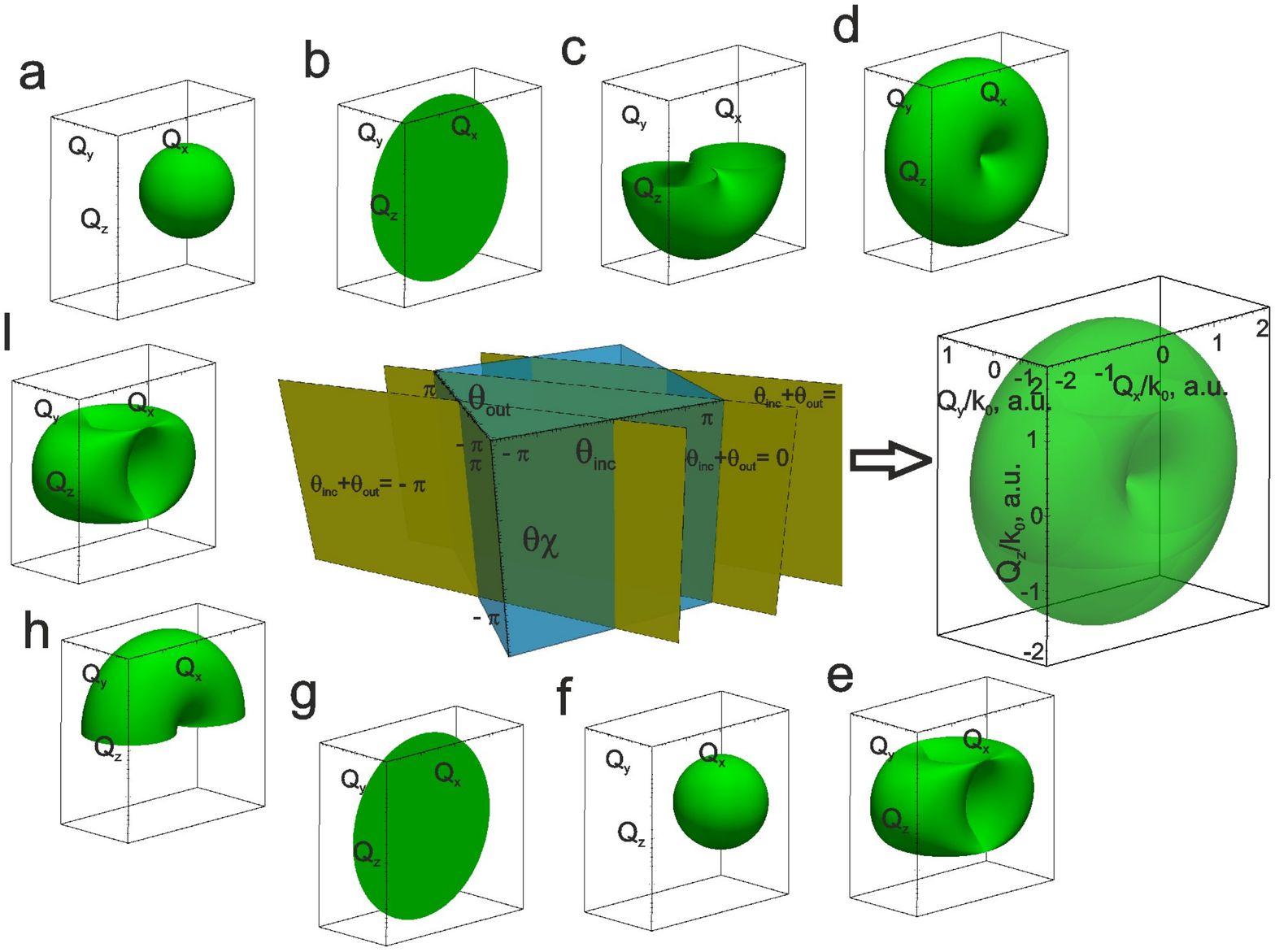}
\end{figure*}

\begin{figure*}
\caption{Accessible reciprocal space at different  ranges of the angles $\theta_{in}$, $\theta_{out}$, $\theta_\chi$.}
\label{Surf3D}
\begin{tabular}{c}
\includegraphics[width=0.8\textwidth]{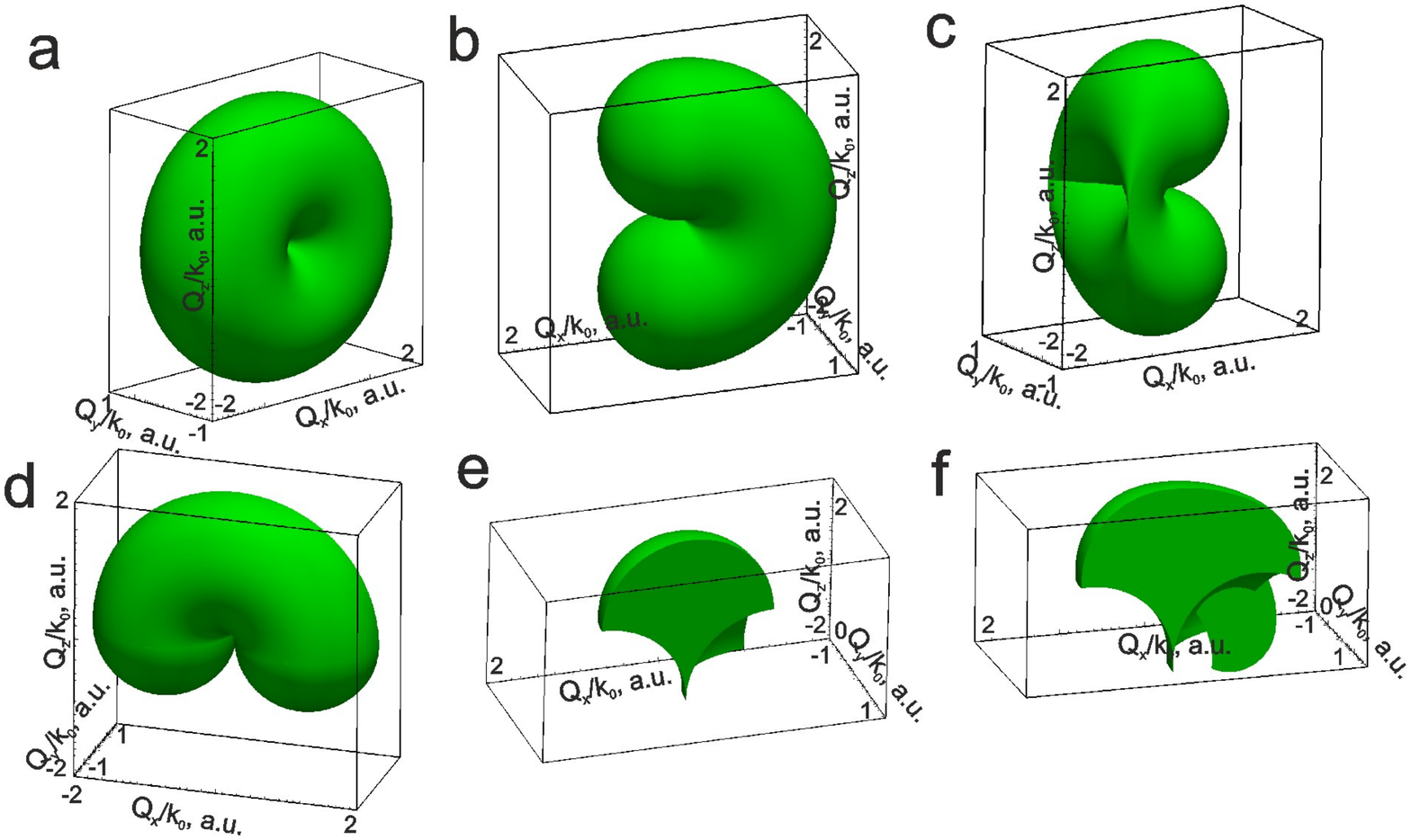} \\
\begin{tabular}{r | c | c | c | c | c | c }
& a & b & c & d & e & f \\ \hline
$\theta_{in, min}$ & $-180^\circ$ & $-90^\circ$ & $-90^\circ$ & $0^\circ$ & $0^\circ$ & $-5^\circ$\\
$\theta_{in, max}$ & $180^\circ$ & $90^\circ$ & $90^\circ$& $180^\circ$ & $90^\circ$ & $95^\circ$\\
$\theta_{out, min}$ & $-180^\circ$ & $-180^\circ$ & $-90^\circ$ & $0^\circ$ & $0^\circ$ & $-5^\circ$ \\
$\theta_{out, max}$ & $180^\circ$ & $180^\circ$ & $90^\circ$ & $180^\circ$ & $90^\circ$ & $120^\circ$\\
$\theta_{\chi, min}$ & $-180^\circ$ & $-180^\circ$ & $-90^\circ$ & $0^\circ$ & $0^\circ$ & $-3^\circ$ \\
$\theta_{\chi, max}$ & $180^\circ$ & $180^\circ$ & $90^\circ$ & $180^\circ$ & $90^\circ$ & $120^\circ$\\
\end{tabular}\\
\end{tabular}
\end{figure*}

The drawing of two parametric surfaces in the case of the vector $\vec{Q}(\theta_{in},\theta_{out})$ is straightforward, however,  in the case of goniometer with the additional axis $\theta_\chi$ becomes a quite complicated task because of the accessible reciprocal space is then described by three parametric function  $\vec{Q}(\theta_{in}, \theta_{out}, \theta_\chi)$. In that case, the drawing of the accessible reciprocal space is reduced to the outer surface mapping $\vec{Q}(\theta_{in},\theta_{out}, \theta_\chi)$. To plot this complex picture, the property of the conformal mappings of a simply connected regions is used, which transfer the boundaries of one region into the boundaries of the second region. The region $(\theta_{in},\theta_{out}, \theta_\chi)$ is quite easy represented, however, the regions of a single-valued correspondence $(\theta_{in},\theta_{out}, \theta_\chi) \longleftrightarrow (Q_x, Q_y, Q_z)$ have to be found. The condition $\left| \vec{Q} \right|' = 0$ defines the part of the outer surface $(Q_x, Q_y, Q_z)$ or the boundary of the regions of a single-valued correspondence, and at fixed $\theta_\chi$  this condition is satisfied when $\theta_{in} + \theta_{out} = n \pi$. Figure~\ref{Surf3DFull} shows the drawings for all surfaces of the angular space $\left\{ \theta_{in} (-\pi, \pi),\theta_{out} (-\pi, \pi), \theta_\chi(-\pi,\pi) \right\}$  and the final shape of the accessible reciprocal space is demonstrated, too. The accessible reciprocal space shapes for the different angle ranges $\theta_{in}$, $\theta_{out}$, $\theta_\chi$ are shown on Fig.~\ref{Surf3D}.

\section{Resolution element}
\label{sec:3}

For the proper analysis of the measured X-ray diffraction data, both the resolution function of the optical system and the contribution of a sample form and size into the intensity of the detected signal have to be taken into account. The accounting of the additional degree of freedom of the detector arm introduces the additional complications \cite{2014BenediktovitchIF}. The accurate calculation of these effects in the measured data is quite tedious, and therefore it is important to estimate the system resolution before the measurements in order to evaluate the computational complexity of analysis or to introduce the correct simplifications into the resolution function. The classical definition of the resolution element is presented, for example, in \cite{pietsch2004}. For the accurate expression of the reciprocal space resolution element it is necessary to add to Eq.~(\ref{formulasSurf3D}) an additional degree of freedom for $\vec{k}_{in}$, that corresponds to a small deviation of the incidence beam in plane $L_xL_y$ due to a linear focal spot of X-ray tube and the absence of the optical limitations of the beam in this direction.

\begin{subequations}
\begin{align}
\vec{k}_{in}^{'} = k_0 \mathbf{\Theta_{in}}\mathbf{\Theta_{\phi}} \vec{L}_x,\\
\mathbf{\Theta_\phi} = \mathbf{R}(\theta_\phi, \vec{L}_z),\\
\vec{k}_{in}^{'} = \{\cos (\theta_i) \cos (\theta_\phi ),\sin (\theta_\phi ),
\sin ( \theta_i )(-\cos (\theta_\phi ))\}, \\
\vec{Q}^{'} = \vec{k}_{out} - \vec{k}_{in}^{'},
\end{align}
\begin{align}
\vec{Q}^{'} = \{\cos (\theta_{out}) \cos (\theta_\chi )-\cos (\theta_{in})
   \cos (\theta_\phi ),
   -\sin (\theta_\chi )-\sin (\theta_\phi ),\nonumber \\
   \sin
   (\theta_{in}) \cos (\theta_\phi )+\sin (\theta_{out}) \cos
   (\theta_\chi )\},
\end{align}
\end{subequations}
where angle $\theta_\phi$ is the angle of a small deviation of the line connecting sample and X-ray source in horizontal plane (see Fig.~\ref{VerticalGonio}(a)).

The function of the resolution element at the fixed reciprocal space point parameterized with the angles $\theta_{in}$, $\theta_\phi$, $\theta_{out}$, $\theta_\chi$, in the case if $\theta_\phi = 0$ has the following form:

\begin{subequations}
\begin{align}
\vec{Q}^{'} = \{\cos (\theta_{out}+\delta \theta_{out}) \cos (\theta_\chi+ \delta \theta_\chi)
-\cos (\theta_{in}+\delta\theta_{in} ) \cos (\delta \theta_\phi ), \nonumber \\
   -\sin (\theta_\chi+ \delta \theta_\chi)-\sin (\delta \theta_\phi ), \\
   \sin(\theta_{in}+\delta \theta_{in} ) \cos (\delta \theta_\phi )+
   \sin (\theta_{out}+\delta \theta_{out}) \cos
   (\theta_\chi + \delta \theta_\chi)\}, \nonumber
\end{align}
\end{subequations}
\noindent where $\delta\theta_{in}$, $\delta \theta_\phi$, $\delta \theta_{out}$, $\delta \theta_\chi$ are the angular parameters of the resolution of the goniometer optical system.

\begin{figure*}
\caption{Accessible reciprocal space: resolution elements shape at different positions of diffractometer. Resolution parameters were taken as $\delta\theta_{in} = 0.001^{\circ}$, $\delta \theta_\phi = 1^{\circ}$, $\delta \theta_{out} = 0.5^{\circ}$, $\delta \theta_\chi = 3^{\circ}$ and they were kept the same at all points.}
\label{ResEl}
\begin{tabular}{c}
\includegraphics[width=0.8\textwidth]{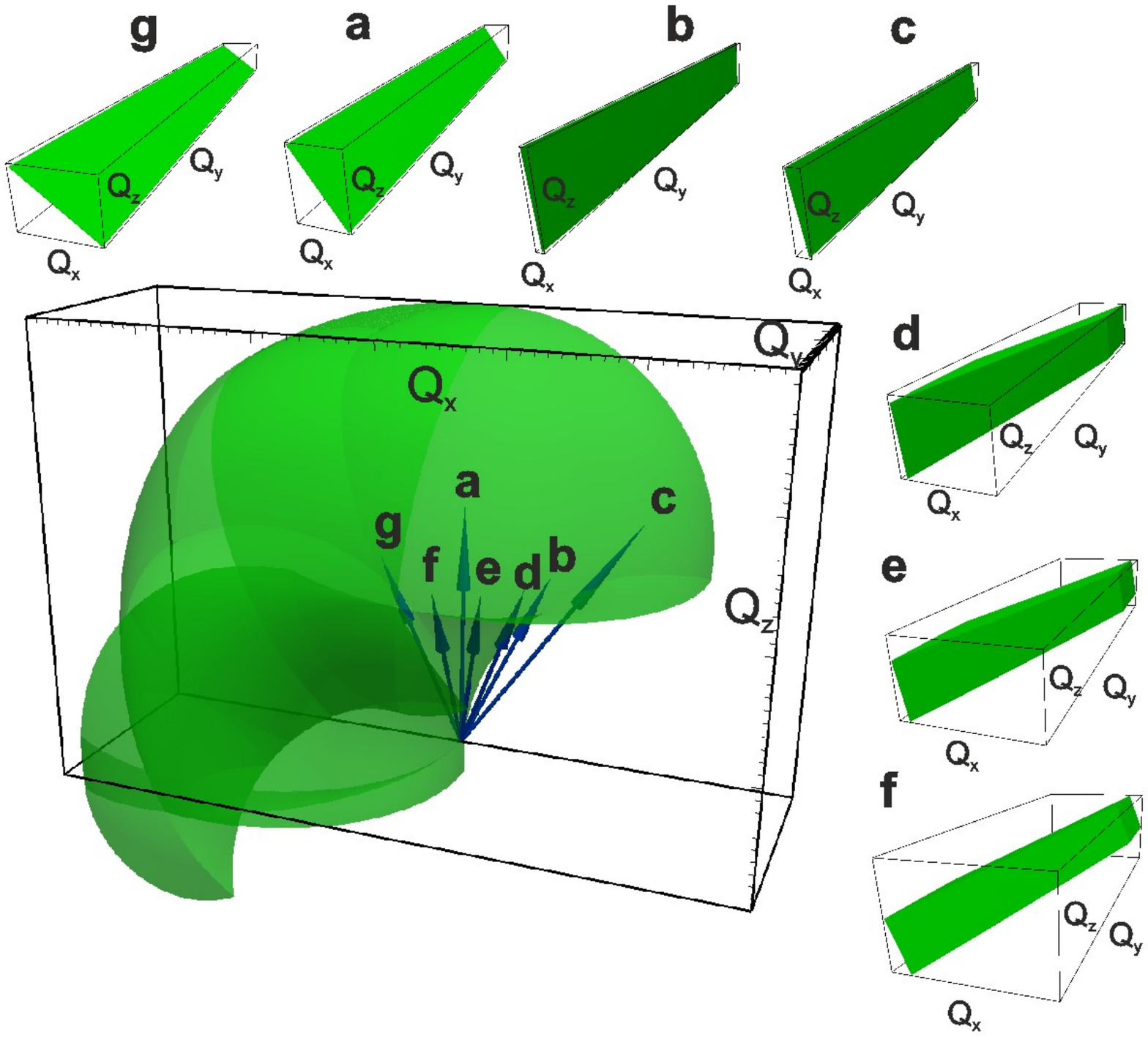} \\
\begin{tabular}{r|l |l |l |l |l |l |l}
& g & a & b & c & d & e & f \\ \hline
$\theta_{in}$ & $2.82^\circ$
 & $34.56^\circ$
 & $53.3^\circ$
 & $79.28^\circ$
 & $50.99^\circ$
 & $43.67^\circ$
 & $31.33^\circ$ \\

$\theta_{out}$ & $53.3^\circ$
 & $34.56^\circ$
 & $2.82^\circ$
 & $8.75^\circ$
 & $4.33^\circ$
 & $9.85^\circ$
 & $21.18^\circ$\\

$\theta_{\chi}$ & $0^\circ$
 & $0^\circ$
 & $0^\circ$
 & $0^\circ$
 & $11.57^\circ$
 & $20.3^\circ$
 & $23.65^\circ$\\
\end{tabular}\\
\end{tabular}
\end{figure*}

Assuming the volume of the detected pixel to be small enough and the uniformity of the resolution element, and using the property of conformal mapping of boundaries for connected regions, we can plot the outer boundaries of the resolution element if each of 24 faces of four-dimensional cube of angular space  $\delta\theta_{in}, \delta \theta_\phi, \delta \theta_{out}, \delta \theta_\chi$ is projected into reciprocal space. The shape and the volume of the resolution element essentially depends on the position in reciprocal space. The typical shapes of the resolution element in different point of reciprocal space are shown on Fig.~\ref{ResEl}.

\section{Non-coplanar HRXRD measurements}
\label{sec:4}

In order to define the range of the angular parameters and to set up the measurement geometry, a set of expected reflections and the accessible reciprocal space have to be combined in a single view. This view helps to select a proper region of measurements in reciprocal space. On the basis of the equations (\ref{inverce_problev_formula}), the angular parameters are calculated which are necessary for the proper positioning of goniometer. The selection of the measurement area and the geometry of measurement logically follow from the estimate for the resolution element.

\begin{figure*}
\caption{Reciprocal space maps for the reflections $(113)$, $(\overline{1}\overline{1}3)$, $(1\overline{1}3)$ of the Ge layer on Si substrate.   }
\label{RSMS}
\begin{tabular}{c}
\includegraphics[width=0.8\textwidth]{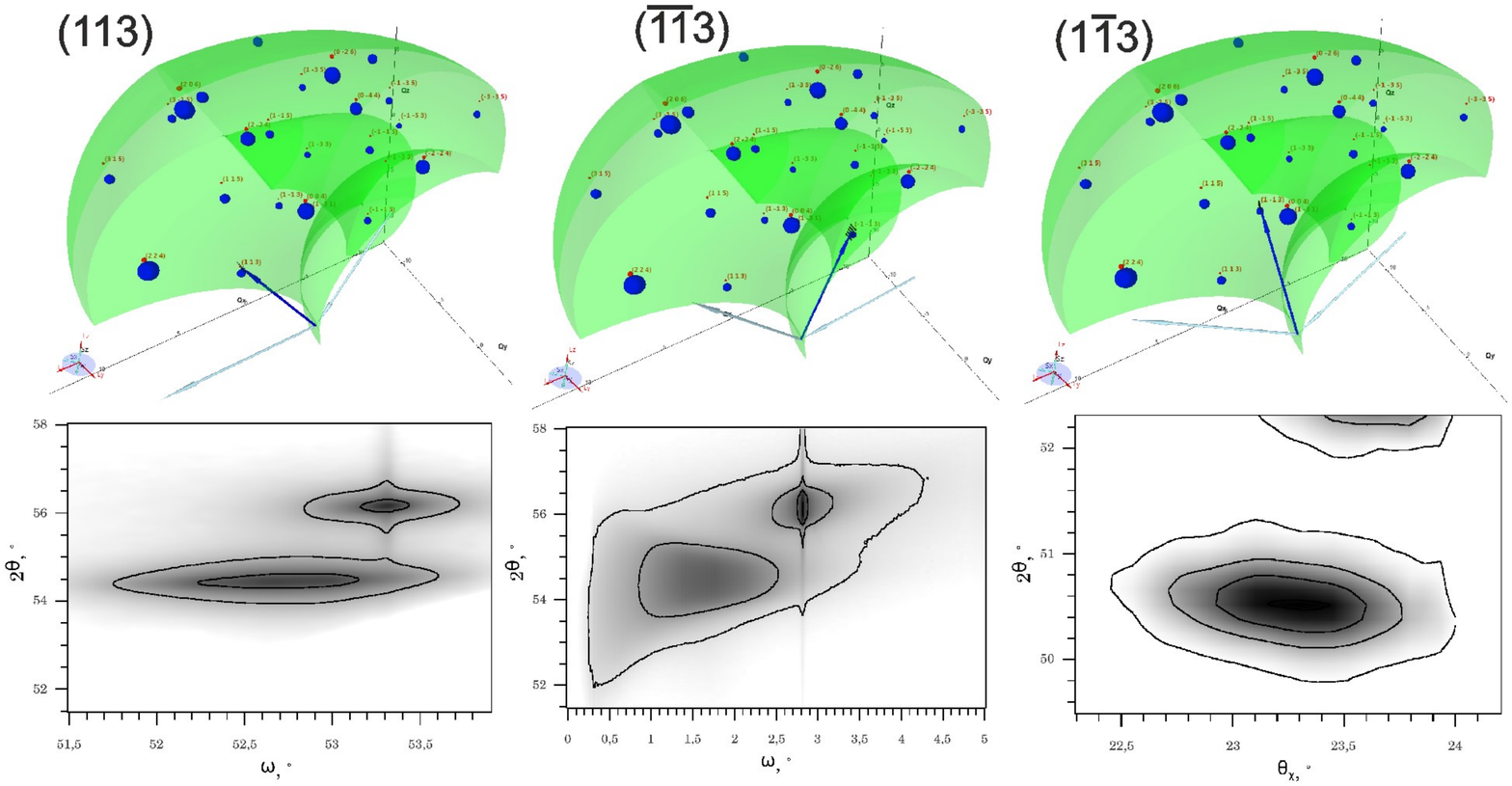} \\
\begin{tabular}{c @{\hspace{5em}} c @{\hspace{5em}} c}
\begin{tabular}{c|c}
 $\theta_{in}$ & $53.3^\circ$ \\
 \hline $\theta_{out}$ & $2.82^\circ$ \\
  \hline $\theta_{\chi}$ & $0^\circ$\\
\end{tabular} &
\begin{tabular}{ c|c }
$\theta_{in}$ & $2.82^\circ$ \\
\hline $\theta_{out}$ & $53.3^\circ$ \\
\hline $\theta_{\chi}$ & $0^\circ$\\
\end{tabular} &
\begin{tabular}{ c|c}
$\theta_{in}$ & $31.33^\circ$ \\
\hline $\theta_{out}$ & $21.18^\circ$ \\
\hline $\theta_{\chi}$ & $23.65^\circ$\\
\end{tabular} \\
\end{tabular} \\

\end{tabular}
\end{figure*}

In order to demonstrate the introduced above scheme of the experiment planning, the reciprocal space mappings have been carried out for the sample consisting of Ge layer on Si substrate and for the following reflections: $(113)$ in a coplanar geometry with large incidence angle, $(\overline{1}\overline{1}3)$ in a coplanar geometry with small incidence angle and $(1\overline{1}3)$ in a non-coplanar geometry. The sample position in space was fixed during all measurements. The $2\theta / \omega$ maps were received for the reflections $(113)$ and $(\overline{1}\overline{1}3)$, and $2\theta / \theta_\chi$ map was measured for the reflection $(1\overline{1}3)$. The results of these measurements are presented in Fig.~\ref{RSMS}, which illustrates the convenience and practicality of proper experiment planning using the above described technique to obtain in an optimal way the measured X-ray intensity from Bragg reflections.

In the most common sense, the expressions for non-coplanar case of X-ray diffraction are more complicated due to non-zero $Y$ component of the diffraction vector. However, a non-coplanar measurement geometry delivers more advantageous information on the sample, which is not accessible in coplanar geometry. For example, the measurements of reciprocal space maps for the reflections $(113)$ and $(\overline{1}\overline{1}3)$ require grazing angles ($\sim 2.8^{\circ}$) for either incidence or exit beam, which makes the analysis quite difficult. In opposite, the $(1\overline{1}3)$ reciprocal space mapping is performed at the incidence and the exit angles, which are far from the grazing angles. The $X$ component of the vector $\vec{Q}$  is close to zero in the case of $(1\overline{1}3)$ reciprocal space map. All the measurements are performed at fixed sample, which is convenient for experimentalist and makes possible to introduce three offsets for the sample alignment correction applied to all measurements.

\section{Conclusions}
\label{sec:5}

The presented approach provides the algorithms for convenient visualization of the accessible reciprocal space in the case when the vertical goniometer with a detector arm possesses two degrees of freedom. The proposed analytical expressions for the angular coordinates in the considered parametrization for the position of the reciprocal lattice vector are used for computer tool assisting the X-ray diffraction experiments. The calculation of the resolution element in the reciprocal space is a part of the presented method and the examples of the resolution elements in different points of the reciprocal space are demonstrated. The advantages of a non-coplanar diffraction with two degrees of freedom provided by detector arm are shortly discussed.


%

\end{document}